\documentclass[aps, prapplied, reprint, superscriptaddress, longbibliography, floatfix]{revtex4-2}
\usepackage[english]{babel}
\usepackage{graphicx}
\usepackage{bm}
\usepackage{mathrsfs}
\usepackage{hyperref}  
\usepackage{amssymb}
\usepackage{xcolor}
\usepackage{amsmath}

\hypersetup{colorlinks=true, citecolor=blue, urlcolor=blue, linkcolor=blue}

\begin{document}
    \title{A micromagnetic model with bidirectional magneto-thermal coupling}
    \author{Peiru Yi}
    \affiliation{State Key Laboratory of Surface Physics and Institute for Nanoelectronic Devices and Quantum Computing, Fudan University, Shanghai 200433, China}
    \affiliation{Department of Physics, Fudan University, Shanghai 200433, China}
    \author{Zian Xia}
    \affiliation{Department of Physics, Fudan University, Shanghai 200433, China}
    \author{Weichao Yu}
    \email{wcyu@fudan.edu.cn}
    \affiliation{State Key Laboratory of Surface Physics and Institute for Nanoelectronic Devices and Quantum Computing, Fudan University, Shanghai 200433, China}
    \affiliation{Zhangjiang Fudan International Innovation Center, Fudan University, Shanghai 201210, China}
    \date{\today}

%\date{\today}

\begin{abstract}
Most conventional micromagnetic frameworks in spin caloritronics rely on a unidirectional coupling approximation, wherein thermal fluctuations drive magnetization dynamics while the feedback of magnetic dissipation onto the thermal reservoir is neglected. Here, we establish a rigorously self-consistent bidirectional magneto-thermal coupling model by integrating the stochastic Landau-Lifshitz-Gilbert (sLLG) equation with a generalized heat transfer equation. In this closed-loop framework, the local temperature acts as a dynamical variable, and the damping-induced dissipation alongside stochastic work dynamically feeds back into the thermal bath as localized heat sources. Utilizing Ito stochastic calculus, we analytically prove that this coupled system strictly obeys the first law of thermodynamics and spontaneously recovers the correct Boltzmann statistics at equilibrium. Spatially resolved micromagnetic simulations further validate the energy exchange mechanism, capturing the finite-bath temperature reduction induced by spatial variation of magnetic moments and the modified density of states under exchange interactions. This bidirectional framework provides a robust microscopic foundation for investigating complex nonequilibrium magneto-thermal dynamics, such as the unidirectional spin-wave heat conveyer effect, paving the way for advanced spin-caloritronic applications.
\end{abstract}


\maketitle

%\tableofcontents

\section{Introduction}

Spin caloritronics, an interdisciplinary field merging spintronics and thermal transport physics, has garnered significant attention due to its fundamental importance in understanding nonequilibrium magnetic phenomena and its potential for developing low-power spintronic devices \cite{UCHIDA2026174195,Bauer2012SpinCaloritronics,Yuan2026PlasmonicAFM,Chumak2015MagnonSpintronics}. The core of this field lies in the intricate coupling among spin transport, heat flow, and magnetization dynamics. Extensive research has been devoted to exploring spin-dependent transport and thermally driven effects, including spin transport phenomena \cite{Kikkawa2023SSE,An2013UnidirectionalSH,Adachi_2013,Xiao2008STTMTJ,XiaoBauer2012STT,flipse2012}, spin pumping \cite{Brataas2022QuantumSpinPumping,Maekawa2022SpinCurrentReview}, and the spin Seebeck effect \cite{PhysRevB.35.4959,PhysRevLett.99.066603,Uchida2008SpinSeebeck}. In these processes, the continuous interplay between the magnetic subsystem and the thermal environment is indispensable, making a comprehensive understanding of magneto-thermal coupling crucial for advancing both theoretical frameworks and practical applications.

To model magnetization dynamics at finite temperatures, the Landau-Lifshitz-Gilbert (LLG) and Landau-Lifshitz-Bloch (LLB) equations serve as the fundamental theoretical approaches. While the LLB equation accounts for longitudinal relaxation and is particularly suitable for describing magnetic behavior near the Curie temperature \cite{PhysRevB.86.104414,Soenjaya2025LLB,tzoufras_dynamics_2015}, the stochastic LLG (sLLG) equation remains the most widely adopted model owing to its computational efficiency and broad experimental applicability in describing transverse damping and precession \cite{Hong2024Damping,Zhang2023Skyrmion}. Consequently, numerous micromagnetic simulation frameworks, such as MuMax3 \cite{vansteenkiste2014mumax3}, OOMMF \cite{donahue1999oommf}, MAGPAR \cite{scholz2003magpar}, Nmag \cite{fischbacher2007nmag}, FastMag \cite{chang2011fastmag}, and Boris \cite{lepadatu2020boris}, have been developed to incorporate finite-temperature effects. However, these conventional approaches inherently rely on a \textit{unidirectional} magneto-thermal coupling framework, which rests on two strict physical premises. First, the relaxation dynamics of the thermal reservoir are assumed to be significantly faster than those of the magnetic system, ensuring the bath remains in thermal equilibrium while the spins undergo coherent LLG dynamics. Second, the heat capacity of the thermal bath is treated as effectively infinite compared to the magnetic subsystem, rendering the local temperature immune to the energy dissipated by spin-wave relaxation. Consequently, the feedback effect of the magnetic subsystem on the thermal environment is fundamentally neglected.

To overcome these limitations, we establish a rigorously self-consistent \textit{bidirectional} magneto-thermal coupling model. The theoretical premise of this generalized framework is twofold. First, while the local thermalization remains faster than magnetic relaxation, the macroscopic heat conduction between different spatial locations is \textit{not} overwhelmingly faster than the propagation of spin waves. Therefore, spatial temperature gradients and heat fluxes must be explicitly resolved via a generalized heat conduction equation, rather than being homogenized into a zero-dimensional lumped-capacitance model. Second, the heat capacity of the local thermal bath is comparable to that of the magnetic subsystem. In such finite-bath regimes, the localized temperature variations induced by magnon relaxation become highly significant and dynamically feed back into the stochastic spin dynamics. By integrating a local temperature evolution equation with the sLLG equation, the temperature is treated as a dynamical variable rather than a fixed external parameter. In this closed-loop framework, the magnetic system not only experiences thermal fluctuations but also dissipates energy back into the heat bath through damping processes. We perform comprehensive analytical derivations and numerical simulations utilizing the Micromagnetics Module \cite{zhang2023frequency,xu2025frequency, hua2024micromagnetic, wang2026effective} fully coupled with the Heat Transfer Module in COMSOL Multiphysics to validate the proposed model. The analytical predictions show excellent agreement with the numerical results, demonstrating that our approach effectively captures the bidirectional energy dynamics and naturally recovers the correct thermodynamic equilibrium, providing a robust theoretical and computational foundation for future investigations in spin caloritronics.

\section{Theoretical Framework of Bidirectional Magneto-Thermal Coupling}

We consider a magnetic system discretized into micromagnetic cells. The local dynamics of the normalized magnetization vector $\mathbf{m}(\mathbf{r}, t) = \mathbf{M}(\mathbf{r}, t)/M_s$, where $\mathbf{M}$ is the local magnetization and $M_s$ is the saturation magnetization (with $|\mathbf{m}|=1$), are governed by the stochastic Landau-Lifshitz-Gilbert (sLLG) equation:

\begin{equation}
\frac{\partial \mathbf{m}}{\partial t} = -\gamma \, \mathbf{m} \times (\mathbf{H}_{\text{eff}} + \mathbf{H}_{\text{th}}) + \alpha \, \mathbf{m} \times \frac{\partial \mathbf{m}}{\partial t}.
\label{LLG}
\end{equation}

Here, $\gamma$ is the gyromagnetic ratio, $\alpha$ is the dimensionless Gilbert damping parameter, and $\mathbf{H}_{\text{th}}(\mathbf{r}, t)$ represents the stochastic thermal field arising from the local thermal reservoir. The deterministic driving force is captured by the generalized effective field $\mathbf{H}_{\text{eff}}$, which naturally accommodates different dimensionalities and magnetic interactions. In a zero-dimensional (0D) macrospin model, $\mathbf{H}_{\text{eff}}$ reduces to a uniform external field $\mathbf{H}_{\text{ext}}$. In a one-dimensional (1D) system incorporating exchange interactions, it takes the form $\mathbf{H}_{\text{eff}} = \mathbf{H}_{\text{ext}} + A \frac{\partial^2 \mathbf{m}}{\partial x^2}$, where $A$ is the exchange interaction parameter with the dimension of $\mathrm{A \cdot m}$. This generalized formalism ensures that the fundamental energy exchange mechanism derived below remains universally applicable.

To establish the local energy exchange channel, we analyze the power dissipation and the work done on the magnetic moment system. Taking the cross product of Eq.~(\ref{LLG}) with $\mathbf{m}$, and utilizing the normalization constraint $\mathbf{m} \cdot \frac{\partial \mathbf{m}}{\partial t} = 0$ alongside standard vector identities, we can isolate the damping term. Subsequently, taking the dot product of the resulting expression with $\frac{\partial \mathbf{m}}{\partial t}$ eliminates the precessional components due to orthogonality. This straightforward algebraic manipulation yields the instantaneous local power balance equation:

\begin{equation}
\gamma \mathbf{H}_{\text{eff}} \cdot \frac{\partial \mathbf{m}}{\partial t} + \gamma \mathbf{H}_{\text{th}} \cdot \frac{\partial \mathbf{m}}{\partial t} = \alpha \left|\frac{\partial \mathbf{m}}{\partial t}\right|^2.
\end{equation}

Multiplying both sides by the characteristic energy scale $\mu_0 M_s / \gamma$ (where $\mu_0$ is the vacuum permeability) translates the dimensionless kinematic equation into a rigorous thermodynamic power density balance:

\begin{equation}
\mu_0 M_s \mathbf{H}_{\text{eff}} \cdot \frac{\partial \mathbf{m}}{\partial t} + \mu_0 M_s \mathbf{H}_{\text{th}} \cdot \frac{\partial \mathbf{m}}{\partial t} = \frac{\alpha \mu_0 M_s}{\gamma} \left|\frac{\partial \mathbf{m}}{\partial t}\right|^2.
\label{power_balance}
\end{equation}

\begin{figure}
	\centering
	\includegraphics[width=0.8\columnwidth]{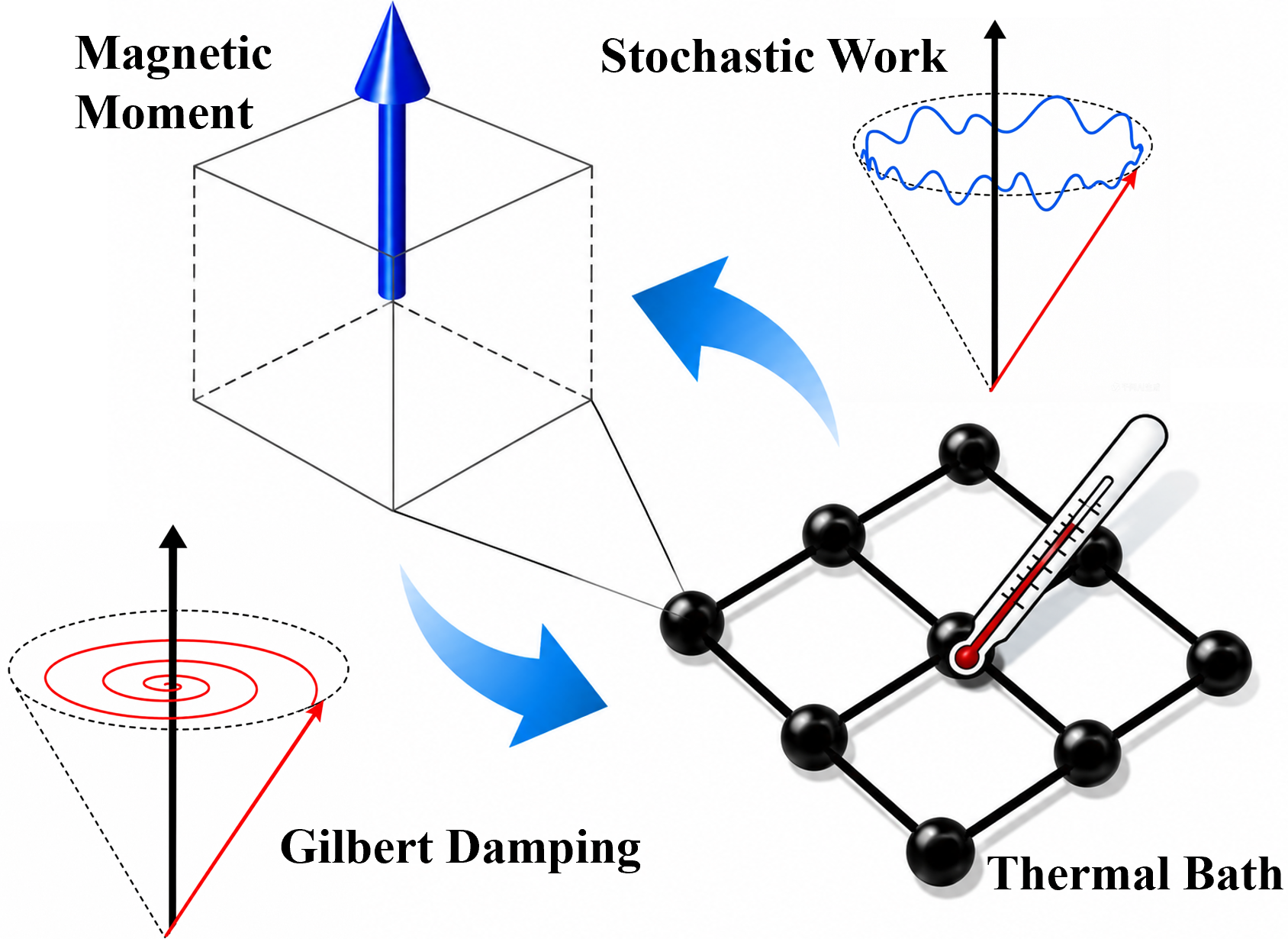}
	\caption{Schematic illustration of the bidirectional magneto-thermal energy exchange mechanism. The magnetic subsystem dissipates energy into the thermal reservoir via Gilbert damping, whereas the thermal reservoir reciprocally transfers energy back to the magnetic system through the work performed by stochastic thermal fields.}
	\label{schematic}
\end{figure}

Eq.~(\ref{power_balance}) reveals the fundamental mechanism of the bidirectional magneto-thermal coupling at the local level. Unlike conventional unidirectional models where damping-induced dissipation is discarded into an infinite heat bath, our framework treats the combined magneto-thermal system as thermodynamically isolated (or locally isolated in the absence of spatial heat diffusion). Consequently, the strictly positive-definite dissipation term on the right-hand side must act as a dynamic, localized volumetric heat source feeding back into the thermal reservoir.

To complete this theoretical framework, we couple the magnetization dynamics with the temperature evolution of the thermal reservoir. Let $T(\mathbf{r}, t)$ denote the local temperature field and $c_V$ be the volumetric heat capacity of the lattice. Based on the energy conservation derived above, the temperature evolution is governed by the generalized heat transfer equation:

\begin{equation}
c_V \frac{\partial T}{\partial t} = \nabla \cdot (\kappa_{\text{th}} \nabla T) + \frac{\alpha \mu_0 M_s}{\gamma} \left|\frac{\partial \mathbf{m}}{\partial t}\right|^2 - \mu_0 M_s \mathbf{H}_{\text{th}} \cdot \frac{\partial \mathbf{m}}{\partial t}.
\label{heat_eq}
\end{equation}

Here, $\kappa_{\text{th}}$ represents the thermal conductivity. This partial differential equation explicitly incorporates spatial heat conduction and the two magneto-thermal source terms illustrated in Fig.~\ref{schematic}: the irreversible Joule-like heating generated by Gilbert damping, and the stochastic work done by the thermal field, which acts as a dynamic heat sink or source. In a 1D scenario, the conduction term reduces to $\kappa_{\text{th}} \frac{\partial^2 T}{\partial x^2}$, describing thermal diffusion along the chain. In a 0D macrospin model, spatial gradients vanish ($\nabla T = 0$), and integrating Eq.~(\ref{heat_eq}) over the cell volume $\Delta V$ perfectly recovers the ordinary differential equation governed by the total heat capacity $C_{\text{bath}} = c_V \Delta V$. 

By solving the sLLG equation [Eq.~(\ref{LLG})] concurrently with this generalized heat transfer equation [Eq.~(\ref{heat_eq})], we establish a fully self-consistent, bidirectional magneto-thermal coupling framework. In this closed loop, the local temperature dictates the amplitude of the stochastic thermal field via the fluctuation-dissipation theorem, while the magnetization dynamics simultaneously govern the temperature evolution through continuous energy exchange, guaranteeing that the coupled system rigorously obeys the first law of thermodynamics and naturally evolves toward the correct equilibrium \cite{widom_gilbert_2010}.

\section{Stochastic Energy Balance and Equilibration Dynamics}

To rigorously verify whether the proposed bidirectional coupling framework naturally recovers thermodynamic equilibrium, we must analyze the energy evolution of the system under stochastic dynamics. Since the sLLG equation is driven by white noise, standard calculus is invalid; instead, we employ It\^o stochastic calculus.

Let the magnetization vector be parameterized in spherical coordinates as $\mathbf{m} = (\sin\theta\cos\phi, \sin\theta\sin\phi, \cos\theta)^T$, and let $\mathbf{\Omega} = (\theta, \phi)^T$ denote its angular coordinates. The sLLG equation [Eq.~(\ref{LLG})] can be recast into a system of coupled stochastic differential equations (SDEs)\cite{194632,Berkov2002FastSwitching}:
\begin{equation}
d\mathbf{\Omega} = \mathbf{A}(\mathbf{\Omega}) dt + \mathbf{B}(\mathbf{\Omega}, T) d\mathbf{W}_t,
\label{sde}
\end{equation}
where $\mathbf{A}$ is the deterministic drift vector, $d\mathbf{W}_t$ is a vector of independent Wiener increments, and $\mathbf{B}(\mathbf{\Omega}, T)$ is the diffusion matrix. Crucially, because the stochastic thermal field satisfies the fluctuation-dissipation theorem ($\mathbf{H}_{\text{th}} \propto \sqrt{T}$), the diffusion matrix $\mathbf{B}$ explicitly depends on the dynamic temperature $T$. In conventional unidirectional models, $T$ is a fixed parameter and $\mathbf{B}$ is constant; however, in our bidirectional framework, $T$ is a dynamical state variable, rendering the system inherently nonlinear.

To track the energy exchange, we apply the multidimensional It\^o lemma to the total magnetic energy $E(\mathbf{\Omega})$ of a single macrospin cell. The non-vanishing quadratic variation of the Wiener process ($(dW_i)^2 = dt$) introduces a crucial second-order correction:
\begin{equation}
dE = \nabla_\Omega E \cdot d\mathbf{\Omega} + \frac{1}{2} \text{Tr}\big(\mathbf{B}^T \mathcal{H}_E \mathbf{B}\big) dt,
\label{ito_E}
\end{equation}
where $\nabla_\Omega=\partial/\partial\Omega$ and $\mathcal{H}_E$ is the Hessian matrix of the energy landscape. The first-order term encompasses the deterministic work and the instantaneous stochastic work done by the thermal field. The second-order It\^o correction term acts as a noise-induced energy drift, representing the continuous net energy injected into the magnetic subsystem by thermal fluctuations. This drift precisely counterbalances the irreversible energy dissipation caused by Gilbert damping.

Under the thermodynamic adiabatic approximation for the 0D macrospin model, the energy lost or gained by the magnetic subsystem must be exactly compensated by the local thermal reservoir. Recalling that the total heat capacity of the discretized cell is $C_{\text{bath}} = c_V \Delta V$, the temperature evolution is governed by the stochastic differential equation:
\begin{equation}
\begin{split}
C_{\text{bath}} dT &= -dE \\
&= -\nabla_\Omega E \cdot \mathbf{B}(\mathbf{\Omega}, T) d\mathbf{W}_t \\
&\quad - \Bigl( \nabla_\Omega E \cdot \mathbf{A} + \frac{1}{2} \text{Tr}\bigl(\mathbf{B}^T \mathcal{H}_E \mathbf{B}\bigr) \Bigr) dt.
\end{split}
\label{sde_T}
\end{equation}

Equation~(\ref{sde_T}) mathematically exposes the core feature of the bidirectional coupling. Because $\mathbf{B} \propto \sqrt{T}$, the temperature evolution equation itself contains multiplicative noise terms proportional to $\sqrt{T} d\mathbf{W}_t$ and deterministic drift terms proportional to $T dt$. This forms a highly nonlinear, self-consistent feedback loop: the Gilbert damping dissipates energy into the heat bath (increasing $T$), which in turn amplifies the stochastic thermal field (increasing $\mathbf{B}$). Conversely, the stochastic field performs work on the magnetic moment system (decreasing $E$), which cools the heat bath via the energy conservation constraint. 

\begin{figure}
	\centering
	\includegraphics[width=1\columnwidth]{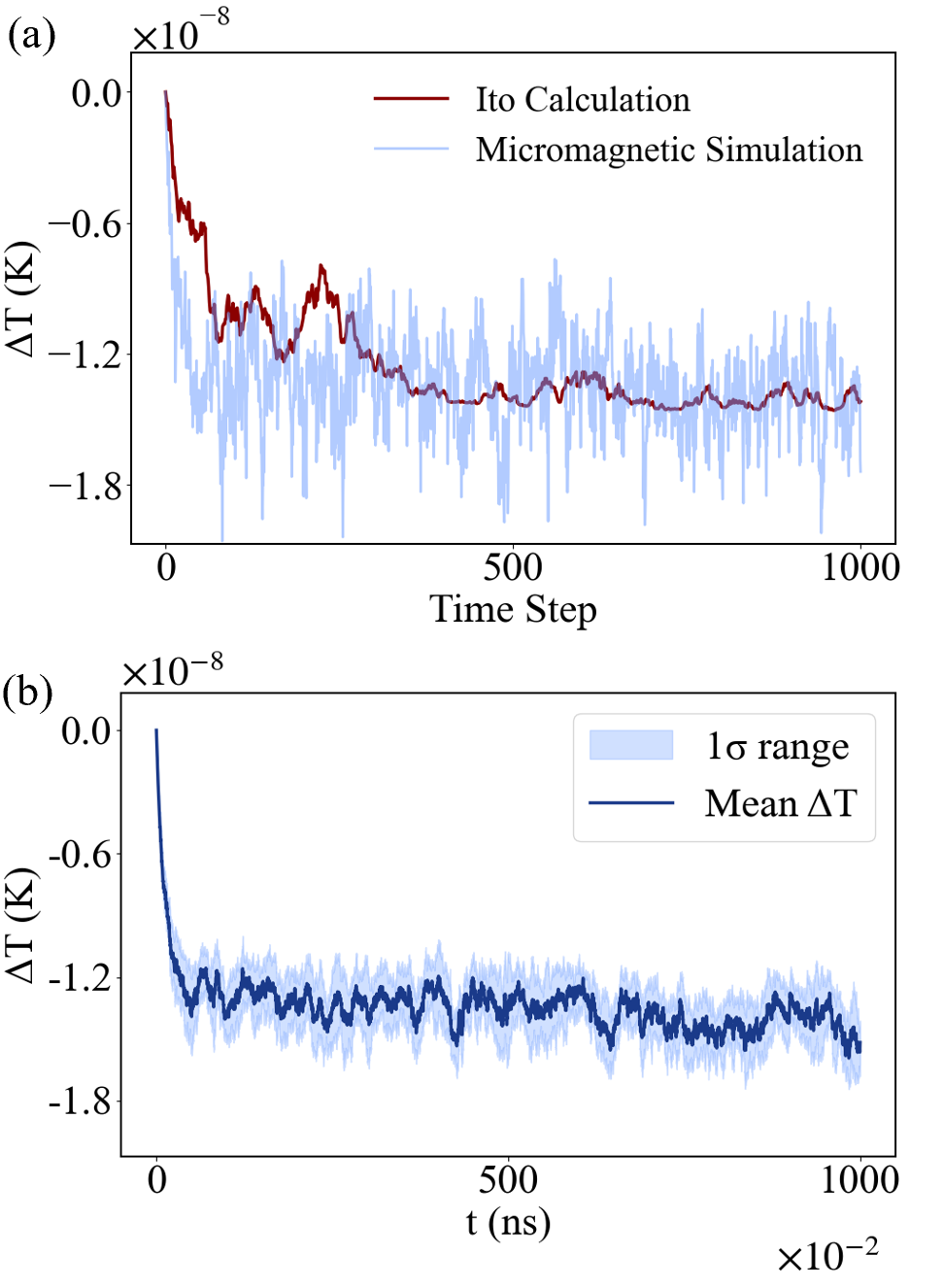}
	\caption{Time evolution of the heat bath temperature under bidirectional coupling. (a) Single-point temperature trajectory simulated using Ito calculation with gyromagnetic ratio $\gamma = 2.21 \times 10^3\,\mathrm{m/(A \cdot s)}$ , macrospin volume of $V = 5 \times 10^{-22}\,\mathrm{m}^3$ and initial temperature $T_0 = 2\,\mathrm{K}$. In micromagnetic simulation, one time step equals $1\times 10^{-2} \mathrm{ns} $. (b) Spatially averaged temperature evolution over 200 micromagnetic cells simulated in COMSOL Multiphysics with cell volume $\Delta V = 5 \times 10^{-22}\,\mathrm{m}^3$. The blue shaded region indicates the $1\sigma$ confidence interval of the temperature distribution across spatial points. Inset: Steady-state temperature variance as a function of various initial temperatures $T_0$.}
	\label{temp}
\end{figure}

It is precisely this nonlinear bidirectional feedback that prevents the system from collapsing into the ground state or diverging to infinity. Instead, it drives the coupled system toward a unique steady state where the energy injection and dissipation balance on average. To validate this, we numerically integrate the fully coupled SDEs for $\mathbf{\Omega}$ and $T$. As shown in Fig.~\ref{temp}, despite the continuous stochastic energy exchanges, the heat bath temperature exhibits a well-defined convergence, confirming the establishment of a steady-state magneto-thermal equilibrium.

 To numerically validate the theoretical predictions shown in Fig.~\ref{temp}(a), the simulated material is yttrium iron garnet (YIG), chosen for its low damping and excellent spin-wave properties \cite{Levati2025}. The intrinsic material parameters are set as follows: saturation magnetization $M_s = 1.94 \times 10^5\,\mathrm{A/m}$, mass density $\rho = 5170\,\mathrm{kg/m^3}$, and specific heat capacity $c_p = 590\,\mathrm{J/(kg \cdot K)}$ \cite{kim_thermal_2017}, yielding a volumetric heat capacity $c_V = \rho c_p \approx 3.05 \times 10^6\,\mathrm{J/(m^3 \cdot K)}$. To ensure statistical convergence within a reasonable computational time, the simulation adopts gyromagnetic ratio $\gamma = 2.21 \times 10^3\,\mathrm{m/(A \cdot s)}$ and macrospin volume of $V = 5 \times 10^{-22}\,\mathrm{m}^3$. The system is initialized at $T_0 = 2\,\mathrm{K}$ with an enhanced damping coefficient $\alpha = 0.05$. The stochastic trajectories are integrated with a time step of $\Delta t = 1 \times 10^{-12}\,\mathrm{s}$.

\begin{figure*}[t!]
	\centering
	\includegraphics[width=1.6\columnwidth]{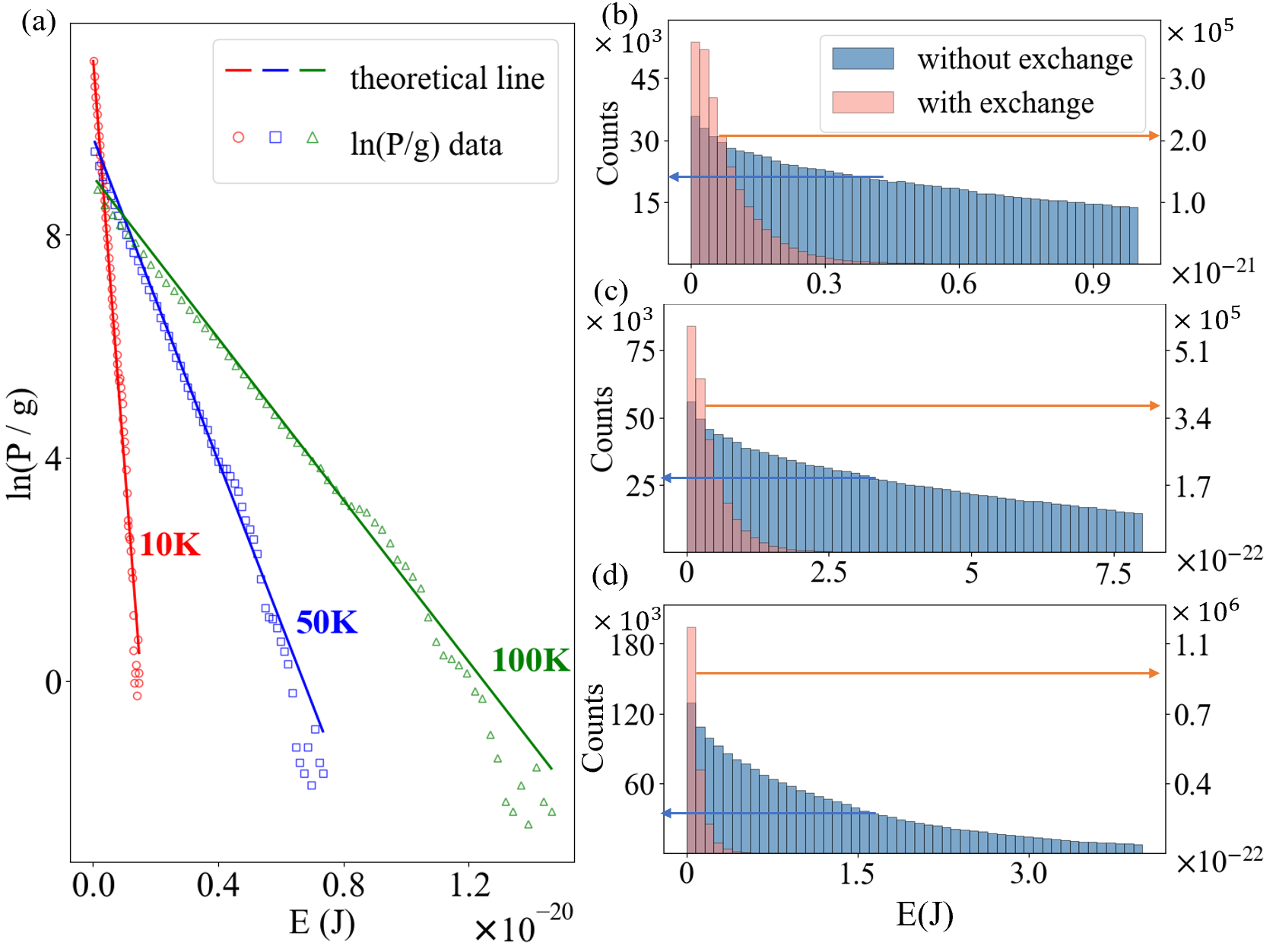}
	\caption{Verification of the Boltzmann statistical distribution using COMSOL Multiphysics. The 1D system consists of 200 cells with volume $\Delta V = 5 \times 10^{-22}\,\mathrm{m}^3$, time step $\Delta t = 1\,\mathrm{ps}$, and damping $\alpha = 0.05$. (a) Semi-logarithmic plot of the numerically sampled $\ln(P/g)$ versus energy $E$, showing strict linear agreement with the theoretical slope $-\beta$. (b)-(d) Statistical histograms of the magnetic moment energy distribution at thermal equilibrium for $T = 100\,\mathrm{K}$, $50\,\mathrm{K}$, and $10\,\mathrm{K}$, respectively. The solid lines represent the analytical prediction $P(E) \propto g(E)e^{-E/k_B T}$.}
	\label{FIG:3}
\end{figure*}

\section{Implementation and Verification via Micromagnetic Simulations}

To rigorously verify the steady-state statistics, we perform spatially resolved micromagnetic simulations using the Micromagnetics Module fully coupled with the Heat Transfer Module in COMSOL Multiphysics. Specifically, we model a one-dimensional magnetic system with a total length of $2000\,\mathrm{nm}$. The system is discretized into 200 equally spaced grid cells, yielding a realistic micromagnetic cell volume of $\Delta V = 5 \times 10^{-22}\,\mathrm{m}^3$. The intrinsic material parameters of YIG remain identical to those defined in Section 3, and an external magnetic field $H_{\text{ext}} = 3.88 \times 10^4\,\mathrm{A/m}$ is applied along the $z$-axis. The exchange interaction parameter is set to $A = 0.328 \times 10^{-10}\,\mathrm{A \cdot m}$. The thermal field satisfies the fluctuation-dissipation theorem \cite{brown_thermal_1963}:
\begin{equation}
\mathbf{H}_{\mathrm{th}}(t) = \sqrt{\frac{2\alpha k_B T}{\gamma \mu_0 M_s \Delta V \Delta t}} \boldsymbol{\eta}(t),
\label{thermal_field}
\end{equation}
where $\boldsymbol{\eta}(t)$ represents a vector of independent Gaussian random variables with zero mean and unit variance generated at each time step \cite{gardiner2009stochastic}. Data from all 200 grid cells are collected over the simulation time for statistical analysis to ensure robust ensemble averaging. To overcome the poor convergence of default solvers under discontinuous white noise, we employ the Generalized Alpha scheme \cite{comsol} with a high-frequency amplification factor of 0.99. This minimizes artificial numerical dissipation, ensuring robust convergence while preserving genuine thermal fluctuations. The simulations use a strict time step $\Delta t = 1 \times 10^{-12}\,\mathrm{s}$ and damping $\alpha = 0.05$.

According to statistical physics, the steady-state distribution of magnetic moments must obey the Boltzmann distribution \cite{Rogel-Salazar01112011}. For a single macrospin under $\mathbf{H}_{\text{ext}}$, the magnetic energy is $E(\theta) = \mu_0 M_s \Delta V H_{\text{ext}} (1-\cos\theta)$, where $\theta$ is the polar angle. Due to the nonlinear dependence of $E$ on $\theta$, the probability density requires the introduction of the density of states $g(E)$, yielding $P(E) \propto g(E)e^{-\beta E}$ with $\beta = (k_B T)^{-1}$. 

By evaluating the phase-space integral over the solid angle, the analytical expression for the density of states under a purely Zeeman coupling is obtained as:
\begin{equation}
g_0(E) = \frac{1}{2 \mu_0 M_s \Delta V H_{\text{ext}}}.
\end{equation}
When spatial exchange interactions are incorporated, the local effective field becomes configuration-dependent, yielding the generalized density of states:
\begin{equation}
g(E) = \frac{1}{2 \mu_0 M_s \Delta V \left( H_{\text{ext}} + A\frac{m_{i+1,z}+m_{i-1,z}-2m_{i,z}}{d^2} \right)}.
\end{equation}
The rigorous derivation of these expressions via the Dirac delta function is detailed in the Appendix.

In Fig.~\ref{FIG:3}, we compare the numerical steady-state histograms extracted from the bidirectional COMSOL simulations at $T = 10, 50,$ and $100\,\mathrm{K}$ with the theoretical prediction $P(E) \propto g(E)e^{-\beta E}$. The excellent agreement confirms that the proposed bidirectional coupling framework naturally drives the 1D system to the correct thermodynamic equilibrium, accurately recovering the Boltzmann statistics without any ad hoc temperature rescaling. Furthermore, a detailed analysis of the distributions in Figs.~\ref{FIG:3}(b)--(d) reveals that the incorporation of exchange interactions significantly alters the energy profile. Specifically, the energy distribution becomes more sharply concentrated in the low-energy regime. Physically, this concentration arises because the ferromagnetic exchange coupling introduces strong spatial correlations between adjacent macrospins, effectively enhancing the local magnetic stiffness. This cooperative effect increases the effective energy barrier for magnetic moment deviations, thereby suppressing large-angle thermal fluctuations and confining the magnetic moments more tightly around the low-energy ground state.

\section{Energy Conservation and Temperature Reduction in a Finite Heat Bath}

We consider an isolated magneto-thermal system where a single macrospin $\mathbf{m}$ is coupled to a finite heat bath with total heat capacity $C_{\text{bath}}$. Initially, the bath is at temperature $T_0$, and the magnetic moment is perfectly aligned with the external field $H_{\mathrm{ext}}$, corresponding to the minimum magnetic energy $E_i=-\mu_0 M_s \Delta V H_{\mathrm{ext}}$. As the system relaxes to thermal equilibrium at a final temperature $T_f$, the magnetic moment absorbs thermal energy from the bath to sustain its fluctuations. This energy transfer manifests directly as a drop in the bath temperature. Since the total system is isolated, energy conservation dictates:
\begin{equation}
C_{\text{bath}}(\langle T_f \rangle -T_0) = E_i-\langle E_f\rangle .
\label{eq:energy_balance}
\end{equation}

To evaluate how much energy the magnetic moment extracts from the bath (i.e., the equilibrium magnetic moment energy $\langle E_f\rangle$), we consider the strong-field regime ($\mu_0 M_s \Delta V H_{\mathrm{ext}} \gg k_B T_f$). The magnetization is confined to small polar angles ($\theta\ll1$), allowing the expansion $\cos\theta \approx 1-\theta^2/2$. The magnetic energy then becomes a simple harmonic potential:
\begin{equation}
E(\theta) = -\mu_0 M_s \Delta V H_{\mathrm{ext}} + \frac{1}{2} \mu_0 M_s \Delta V H_{\mathrm{ext}}\,\theta^2.
\end{equation}
The corresponding partition function reduces to a standard Gaussian integral:
\begin{equation}
Z \propto \int_0^{\infty} \theta \exp\!\left(-\frac{\beta \mu_0 M_s \Delta V H_{\mathrm{ext}}}{2}\theta^2\right) d\theta \propto \frac{1}{\beta},
\end{equation}
where $\beta=(k_B \langle T_f \rangle)^{-1}$. 

From standard statistical mechanics, the equilibrium average energy is:
\begin{equation}
\langle E_f\rangle = -\frac{\partial}{\partial\beta}\ln Z = -\mu_0 M_s \Delta V H_{\mathrm{ext}} + \frac{N_{\perp}}{2} k_B \langle T_f \rangle ,
\label{eq:final_energy}
\end{equation}
where $N_{\perp}=2$ represents the two transverse degrees of freedom of the magnetic moment. 

Substituting Eq.~(\ref{eq:final_energy}) into the energy conservation relation Eq.~(\ref{eq:energy_balance}) gives:
\begin{equation}
C_{\text{bath}}(\langle T_f\rangle -T_0) = -\frac{N_{\perp}}{2}k_B \langle T_f \rangle .
\end{equation}
In the macroscopic limit where $C_{\text{bath}}\gg k_B$, the equilibrium temperature reduction of the bath is:
\begin{equation}
\Delta T = \langle T_f \rangle
-T_0 \approx -\frac{N_{\perp} k_B T_0}{2 C_{\text{bath}}}.
\label{tempchange}
\end{equation}

\begin{figure}
	\centering
	\includegraphics[width=1\columnwidth]{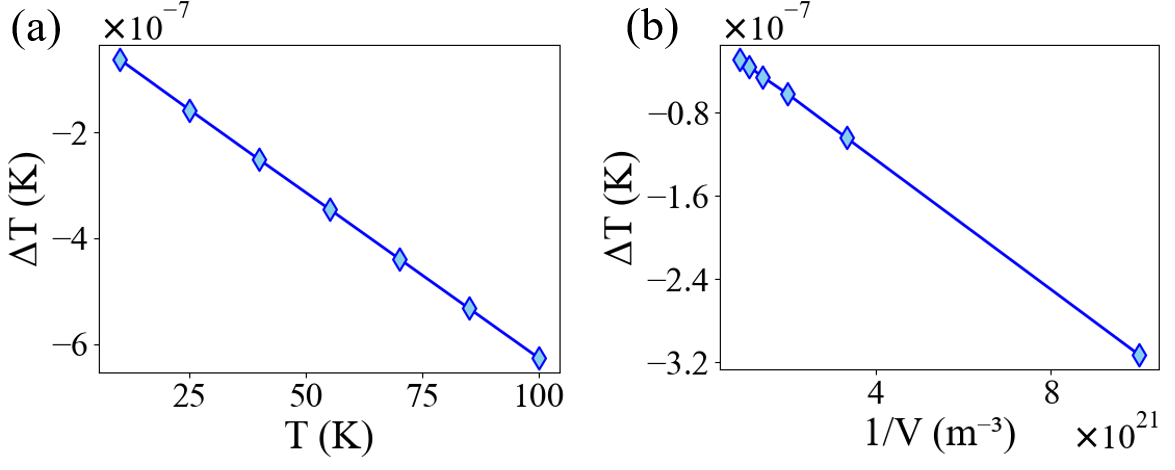}
	\caption{Equilibrium temperature variation of the finite heat bath. (a) Temperature change $\Delta T$ versus initial temperature $T_0$. (b) Temperature change $\Delta T$ versus the macrospin volume $\Delta V$. The simulation parameters are $\Delta V = 5 \times 10^{-22}~\mathrm{m}^3$, $T_0 = 10~\mathrm{K}$, $\Delta t = 1~\mathrm{ps}$, and $\alpha = 5 \times 10^{-3}$.}
	\label{temp_change}
\end{figure}

This result demonstrates that the temperature drop $\Delta T$ is independent of the external field magnitude and the Gilbert damping. Physically, it perfectly quantifies how much heat the magnetic subsystem extracts from the reservoir to establish thermal equilibrium, which is determined solely by the thermal energy scale of the magnetic moment's transverse fluctuations ($\frac{N_{\perp}}{2} k_B T_0$) relative to the bath's heat capacity. As shown in Fig.~\ref{temp_change}, our bidirectional simulations accurately capture this energy-exchange process, confirming that $\Delta T$ is proportional to the initial temperature and inversely proportional to the volume, exactly as predicted by Eq.~(\ref{tempchange}).

\section{Conclusion and Discussion}

In this work, we have established a bidirectional magneto-thermal coupling model by integrating the stochastic Landau--Lifshitz--Gilbert equation with a generalized heat transfer equation. Unlike conventional unidirectional approaches, our framework rigorously captures the mutual energy exchange: thermal fluctuations drive the magnetic subsystem, while damping-induced dissipation and stochastic work dynamically feed back into the local thermal reservoir. Using It\^o stochastic calculus and spatially resolved COMSOL simulations, we demonstrated that this closed-loop system naturally satisfies the first law of thermodynamics and spontaneously recovers the correct Boltzmann statistics at equilibrium.

While the current analytical treatment relies on the low-temperature and strong-field harmonic approximations, the theoretical framework is inherently general. Complex magnetic interactions, such as magnetocrystalline anisotropy, dipolar coupling, and the Dzyaloshinskii--Moriya interaction, can be straightforwardly incorporated via the generalized effective field. On the other hand, the validity of this model is inherently bounded by specific physical regimes. The model is highly applicable to mesoscopic systems where local thermal equilibrium is established much faster than the magnetization dynamics, yet the macroscopic heat diffusion is comparable to the spin-wave propagation. However, the framework encounters fundamental limitations when these underlying assumptions are violated. Specifically, the single-temperature approximation fails under ultrafast excitations where electron, phonon, and spin subsystems strongly decouple. Furthermore, Fourier's macroscopic heat conduction breaks down in ultra-scaled nanostructures exhibiting ballistic thermal transport. Finally, at cryogenic temperatures, our classical stochastic approach necessitates quantum corrections as quantum fluctuations dominate.

A critical aspect of this model is its direct experimental relevance. Although the local temperature variations $\Delta T$ induced by microscopic magneto-thermal exchange are exceptionally small, they are readily accessible via state-of-the-art lock-in thermography (LIT) \cite{kainuma_local_2021}. Advanced LIT techniques now offer an unprecedented temperature resolution of 0.1 mK, enabling the detection of ultra-weak, spin-dependent heat signals with high throughput and spatial precision \cite{SciChinaPress_2024}. This extreme sensitivity has recently facilitated in-plane thermophysical mapping of nanoscale-thick films \cite{Alasli_2025_APL} and the real-space imaging of hidden magnetic textures, such as magnetic octupole domains in non-collinear antiferromagnets \cite{Wang_2024_NSR}. Given these breakthrough capabilities in resolving localized heat sources, the dynamic thermal feedback and asymmetric temperature distributions predicted by our bidirectional model can be directly probed and validated in future spin-caloritronic experiments.

Looking forward, the proposed framework provides a robust foundation for investigating complex nonequilibrium magneto-thermal dynamics. A particularly promising application is the microscopic modeling of the unidirectional spin-wave heat conveyer effect, where the nonreciprocal Damon-Eshbach surface spin waves induce a directional heat drift that competes with isotropic phonon diffusion, resulting in asymmetric local temperature distributions \cite{An2013UnidirectionalSH,10.1063/1.4908019}. Our bidirectional model, capable of resolving local heat emission and spin-wave damping at the micromagnetic cell level, is ideally suited to simulate these intricate heat transfer dynamics. Furthermore, this framework can be readily extended to facilitate deeper investigations into the spin Seebeck effect \cite{Uchida2010SpinSeebeckInsulator,Adachi_2013,Xu2024EnhancingSpinPumping}, the Thomson effect \cite{Takahagi2025TransverseThomson}, and other emerging spin caloritronic phenomena \cite{An2013UnidirectionalSH,Ken-ichiUCHIDA2021PJA9702B-02,Shigematsu2018SpinWaveTemperatureGradient,Hirai2025MagnonThermal,Wid2016UnidirectionalHeatConveyer,Schlitz2025DirectionalMagnon}.

\section*{Acknowledgements}

This work was supported by National Key Research Program of China (Grant No. 2022YFA1403300), the Innovation Program for Quantum Science and Technology (Grant No. 2024ZD0300103) and National Natural Science Foundation of China (Grant No. 12574112 and 12204107). The authors thank Jiang Xiao for inspiring discussions.

%\bibliographystyle{unsrt}

%\bibliography{ref}

\appendix

\section{Appendix}

\subsection{Projection of the sLLG Equation into Spherical Coordinates}
To facilitate numerical integration and energy tracking, it is highly advantageous to project the sLLG equation from the Cartesian basis onto the local spherical coordinate basis. Let the normalized magnetization vector be parameterized as $\mathbf{m} = (\sin\theta\cos\phi, \sin\theta\sin\phi, \cos\theta)^T$. Using the mathematically equivalent Landau-Lifshitz (LL) form of the sLLG equation, the dynamics are governed by:
\begin{equation}
d\mathbf{m} = -\frac{\gamma}{1+\alpha^2} \mathbf{m} \times \mathbf{H}_{\text{tot}} dt - \frac{\alpha\gamma}{1+\alpha^2} \mathbf{m} \times (\mathbf{m} \times \mathbf{H}_{\text{tot}}) dt,
\end{equation}
where $\mathbf{H}_{\text{tot}} = \mathbf{H}_{\text{eff}} + \mathbf{H}_{\text{th}}$ is the total effective field. By applying the chain rule and projecting the Cartesian differentials onto the angular basis vectors $\hat{e}_\theta$ and $\hat{e}_\phi$, we obtain the explicit stochastic differential equations for the polar angle $\theta$ and azimuthal angle $\phi$. 

Defining the thermal fluctuation amplitude coefficient as $\kappa = \sqrt{2k_B / (\gamma\mu_0M_s\Delta V)}$, the angular evolution equations are explicitly given by:
\begin{equation}
\begin{aligned}
d\theta
&=
\frac{\gamma}{1+\alpha^2}
\Bigg[
\Big(
(H_y+H_x\alpha\cos\theta)\cos\phi
-
H_z\alpha\sin\theta
\\
&\qquad\qquad
+
(-H_x+H_y\alpha\cos\theta)\sin\phi
\Big)dt
\\
&\qquad
+
\sqrt{T\alpha}\kappa
\Big(
-\alpha\sin\theta\, dW_z
\\
&\qquad\qquad
+
(\alpha\cos\theta\cos\phi-\sin\phi)\,dW_x
\\
&\qquad\qquad
+
(\cos\phi+\alpha\cos\theta\sin\phi)\,dW_y
\Big)
\Bigg],
\end{aligned}
\label{eq:dtheta}
\end{equation}

\begin{equation}
\begin{aligned}
d\phi
&=
\frac{\gamma}{1+\alpha^2}
\Bigg[
\Big(
H_z
+
\cos\phi
(-H_x\cot\theta+H_y\alpha\csc\theta)
\\
&\qquad\qquad
-
H_y\cot\theta\sin\phi
-
H_x\alpha\csc\theta\sin\phi
\Big)dt
\\
&\qquad
+
\sqrt{T\alpha}\kappa
\Big(
dW_z
\\
&\qquad\qquad
+
(\alpha\cos\phi\csc\theta-\cot\theta\sin\phi)\,dW_y
\\
&\qquad\qquad
-
(\cos\phi\cot\theta+\alpha\csc\theta\sin\phi)\,dW_x
\Big)
\Bigg],
\end{aligned}
\label{eq:dphi}
\end{equation}
where $H_x, H_y, H_z$ are the Cartesian components of the deterministic effective field $\mathbf{H}_{\text{eff}}$, and $dW_x, dW_y, dW_z$ are independent Wiener increments satisfying $(dW_i)^2 = dt$ and $dW_i dW_j = 0$ (for $i \neq j$). From Eqs.~(\ref{eq:dtheta}) and (\ref{eq:dphi}), the deterministic drift vector $\mathbf{A}$ and the $2 \times 3$ diffusion matrix $\mathbf{B}$ defined in the main text can be directly extracted for numerical implementation.

\subsection{Explicit Expansion of the It\^o Energy Correction}
In the main text, the noise-induced energy drift is compactly expressed using the trace of the Hessian matrix: $\frac{1}{2} \text{Tr}(\mathbf{B}^T \mathcal{H}_E \mathbf{B}) dt$. For computational reproducibility, it is necessary to expand this abstract matrix operation into explicit scalar partial derivatives. 

Let the magnetic energy be $E(\theta, \phi)$. According to the multidimensional It\^o lemma, the second-order differential $dE^{(2)}$ is given by:
\begin{equation}
dE^{(2)} = \frac{1}{2} \frac{\partial^2 E}{\partial \theta^2} (d\theta)^2 + \frac{1}{2} \frac{\partial^2 E}{\partial \phi^2} (d\phi)^2 + \frac{\partial^2 E}{\partial \theta \partial \phi} d\theta d\phi.
\label{eq:ito_expansion}
\end{equation}
The quadratic variations of the angular coordinates are determined entirely by the row vectors of the diffusion matrix, $\mathbf{B}_\theta$ and $\mathbf{B}_\phi$:
\begin{equation}
(d\theta)^2 = \|\mathbf{B}_\theta\|^2 dt, \quad (d\phi)^2 = \|\mathbf{B}_\phi\|^2 dt, \quad d\theta d\phi = (\mathbf{B}_\theta \cdot \mathbf{B}_\phi) dt.
\end{equation}
Substituting these relations into Eq.~(\ref{eq:ito_expansion}) yields the exact scalar correction term that must be added to the deterministic energy derivative in any numerical integration scheme (e.g., Euler-Maruyama or Heun methods) to preserve the correct Boltzmann statistics at equilibrium.

\subsection{Derivation of the Density of States with Exchange Interaction}
To analytically verify the Boltzmann distribution, the density of states $g(E)$ must be evaluated. Choosing the local polar axis ($\theta=0$) to align with the effective field direction, the magnetic energy is $E = \mu_0 M_s \Delta V H_{\text{eff}} (1-\cos\theta)$. The density of states is defined via the Dirac delta function over the solid angle:
\begin{equation}
g(E) = \frac{1}{4\pi} \int_0^\pi \sin\theta \, d\theta \int_0^{2\pi} d\phi \, \delta\bigl[E - \mu_0 M_s \Delta V H_{\text{eff}}(1-\cos\theta)\bigr].
\end{equation}
Integrating over $\phi$ yields a factor of $2\pi$. Changing the integration variable to $u = 1-\cos\theta$ (where $du = \sin\theta d\theta$), the integral simplifies to:
\begin{equation}
g(E) = \frac{1}{2} \int_0^2 du \, \delta\bigl[E - \mu_0 M_s \Delta V H_{\text{eff}} u\bigr].
\end{equation}
Using the scaling property of the Dirac delta function, $\delta(ax) = \frac{1}{|a|}\delta(x)$, we obtain the analytical expression for the density of states:
\begin{equation}
g(E) = \frac{1}{2 \mu_0 M_s \Delta V H_{\text{eff}}}.
\end{equation}
Note that for a purely Zeeman coupling, $H_{\text{eff}} = H_{\text{ext}}$, and $g(E)$ is a constant, leading to the standard exponential Boltzmann distribution. However, when complex micromagnetic interactions are present, the effective field becomes configuration-dependent. For instance, considering a simplified one-dimensional exchange interaction, the effective field is modified to:
\begin{equation}
H_{\text{eff}} = H_{\text{ext}} + A\frac{m_{i+1,z}+m_{i-1,z}-2m_{i,z}}{d^2},
\end{equation}
where $A$ is the exchange interaction parameter and $d$ is the lattice spacing. By substituting this local $H_{\text{eff}}$ into the derived $g(E)$, the theoretical probability distribution $P(E) \propto g(E)e^{-\beta E}$ can be accurately constructed and compared against numerical histograms, ensuring the validity of the bidirectional model even in the presence of strong spatial correlations.
\end{document}